\documentclass[]{spie}  

\usepackage{amsmath,amsfonts,amssymb}
\usepackage{graphicx}
\usepackage[colorlinks=true, allcolors=blue]{hyperref}

\title{Digging dark holes on-sky with the Self-Coherent Camera:\\Preliminary results}

\author[a]{Elena Tonucci}
\author[a,b]{Sebastiaan Y. Haffert}
\author[b]{Jared R. Males}
\author[b]{Laird M. Close}
\author[b]{Kyle van Gorkom}
\author[b,c,d,e]{Olivier Guyon}
\author[f]{Alexander D. Hedglen}
\author[b]{Parker T. Johnson}
\author[b]{Maggie Y. Kautz}
\author[c]{Jay K. Kueny}
\author[b]{Jialin Li}
\author[c]{Joshua Liberman}
\author[g]{Joseph D. Long}
\author[c]{Jennifer Lumbres}
\author[a]{Matthijs Mars}
\author[c]{Eden A. McEwen}
\author[h]{Avalon McLeod}
\author[a]{María Eugenia Redondo González}
\author[i]{Lauren Schatz}
\author[c]{Katie Twitchell}
\affil[a]{Leiden Observatory, Leiden University, PO Box 9513, 2300 RA, Leiden, The Netherlands}
\affil[b]{Steward Observatory, The University of Arizona, 933 North Cherry Avenue, Tucson, Arizona}
\affil[c]{Wyant College of Optical Sciences, The University of Arizona, 1630 E University Blvd, Tucson, Arizona}
\affil[d]{Subaru Telescope, National Observatory of Japan, National Institutes of Natural Sciences, 650 N. A'ohoku Place, Hilo, Hawai'i}
\affil[e]{Astrobiology Center, National Institutes of Natural Sciences, 2-21-1 Osawa, Mitaka, Tokyo, Japan}
\affil[f]{Northrop Grumman Corporation, 600 South Hicks Road, Rolling Meadows, Illinois}
\affil[g]{Center for Computational Astrophysics, Flatiron Institute, 162 5th Avenue, New York, New York}
\affil[h]{Draper Laboratory, 555 Technology Square, Cambridge, Massachusetts}
\affil[i]{Starfire Optical Range, Kirtland Air Force Base, Albuquerque, New Mexico}

\authorinfo{Further author information: (Send correspondence to E.T.)\\E.T.: E-mail: tonucci@strw.leidenuniv.nl}

\begin{document} 
\maketitle

\begin{abstract}
Current high-contrast imaging instruments are limited by wavefront errors originating from non-common path aberrations (NCPAs) due, for example, to manufacturing errors in the optics and temperature drifts in the system. These create quasi-static speckles in the final science image that are difficult to distinguish from companions. Therefore, focal plane wavefront sensing and control is needed to suppress speckles. The Self-Coherent Camera (SCC) is a wavefront sensor that allows us to estimate the stellar complex speckle field. In the Fast Atmospheric SCC Technique (FAST), the on-axis starlight hits a coronagraphic focal plane phase mask and is diffracted outside the Lyot stop where it is spatially filtered by a pinhole to create a reference beam. The reference beam and the leaked starlight are recombined on the science plane, creating interference fringes that do not affect the companion, because of incoherence. The focal plane mask was manufactured in-house at Leiden University with Nanoscribe, a micro-3D-printer that uses two-photon polymerization to achieve sub-micron precision in height. We present preliminary results of the first on-sky closed-loop SCC demonstration with the Magellan Adaptive Optics eXtreme (MagAO-X) instrument on the 6.5-meter Magellan Clay telescope at Las Campanas Observatory, Chile. We achieve a 1-$\sigma$ raw contrast improvement of a factor 10 in the desired dark hole region with FAST closed-loop control. In the future, we will show observations of stars with companions and use the SCC in post-processing as a Coherent Differential Imaging (CDI) technique to enhance the contrast even further.
\end{abstract}


\section{INTRODUCTION}\label{sec:intro} 
High-contrast imaging (HCI) is one of the most promising techniques to characterize rocky exoplanets with future observatories\cite{ELT,GMT} and possibly discover life outside of our Solar System\cite{HWO_biosignatures}. To reach this objective, HCI technology must be advanced on two fronts. First, on the coronagraphic side, we need to develop coronagraphs with high throughput and a small inner-working angle (IWA)\cite{coro_review,Tonucci_PIAACMC}. Second, we must perform active focal plane wavefront sensing and control to remove NCPAs, all those errors introduced by optical path differences between the wavefront sensing path and the coronagraphic path\cite{common_path_WFS}. NCPAs must be corrected because they make light leak through the coronagraph, which creates random interference resulting in speckles on the image plane. These have a size of about $\lambda$/D, mimicking companions\cite{speckles}, and the ones with lifetimes from seconds to hours are particularly difficult to calibrate or average out. These speckles are called ``quasi-static". 

Removing quasi-static speckles requires us to sense the electric field on the coronagraphic focal plane and apply a correction upstream. Focal plane wavefront sensing and control can both be performed with different techniques. A usual way to perform the control is by actuating a deformable mirror (DM). A DM shape that creates the measured speckle field is applied, but with opposite sign, allowing us to remove such speckles through destructive interference, thus improving the contrast. This is usually done only within a specified region called ``dark hole". This speckle nulling technique is called Electric Field Conjugation\cite{EFC}. It can either be model-based, or empirical, such as implicit EFC (iEFC\cite{iEFC,iEFC_onsky}). It is usual to perform the wavefront sensing with pair-wise probing (PWP\cite{PWP}). However, PWP requires four measurements for the sensing, to reconstruct amplitude and phase of the real and the imaginary part of the electric field. The Self-Coherent Camera (SCC\cite{SCC,SCC_FPWFS_CDI}) instead, allows us to sense the wavefront with a single measurement, drastically decreasing the time required for the sensing. On the Lyot plane of the coronagraph, a pinhole is placed at an adequate distance. Some on-axis starlight diffracted by a diffractive coronagraphic mask (usually an opaque Lyot mask) falls outside of the Lyot stop and goes through the pinhole, creating a spatially filtered reference beam. This beam coherently interferes with residual starlight leaking through the coronagraph. This creates fringes on the image plane that only affect starlight, because off-axis planetary light does not get diffracted towards the pinhole and is incoherent with starlight, so, it does not take part in the interference. The stellar complex field can then be estimated independently from the companion, from a single measurement and through a process called ``side-band extraction", detailed in section \ref{sec:method}.

Different ``flavours" of the SCC have been introduced in time\cite{multiple_pinholes,SCC_Martinez,SCC_Sebastiaan,SCC_Josh} to solve some problems like the large pinhole distance required, or the low throughput in the pinhole. Specifically, to perform the wavefront sensing in real time with ground-based telescopes that are usually limited by the amounts of photons received, we require that as much light as possible falls on the pinhole. To solve this, the Fast Atmospheric SCC Technique (FAST\cite{FAST}) introduced a phase mask on the focal plane of the coronagraph with a wedge. The wedge angle is designed to diffract light at the correct deflection angle and fall on the pinhole. This creates more signal in the reference beam than with a general opaque Lyot mask.

The SCC and FAST have been demonstrated in laboratory conditions in different studies, and some attempts have been made to actively run them in closed-loop on-sky, without success. At the THD2 bench of the Paris Observatory\cite{THD2}, Singh et al. 2019\cite{SCC_lab_THD2} demonstrated a 1-$\sigma$ raw contrast below 10\textsuperscript{-6} in the region 5-12 $\lambda$/D at 783.25 nm. At the Santa Cruz Extreme AO Lab\cite{SEAL}, Gerard et al. 2021\cite{SCC_lab_Gerard} reached a 5-$\sigma$ contrasts within the dark hole of about 5$\times$10\textsuperscript{-4}. 
On the Stellar Double Coronagraph\cite{SDC} instrument at Palomar Observatory, Galicher et al. 2019\cite{Palomar_sky} showed an improvement of a factor 4 to 20 between 1.5 and 5 $\lambda$/D with the internal source. They were unable to perform active closed-loop control on-sky, but passively applied the same correction than in the lab, which still provided an improvement of a factor 5 between 2 and 4 $\lambda$/D. Another group of researchers is currently working on demonstrating FAST in closed-loop on-sky with their instrument called Subaru Pathfinder Instrument for Detecting Exoplanets and Retrieving Spectra (SPIDERS\cite{SPIDERS}), hosted with SCExAO\cite{SCExAO} at the Subaru telescope.

We present preliminary results of the first closed-loop on-sky SCC demonstration with FAST, that we performed in April 2025 with MagAO-X on the 6.5-meter Magellan Clay Telescope at Las Campanas Observatory, in Chile. In section \ref{sec:method}, we detail the layout of the system and explain the wavefront sensing and control process with the SCC. In section \ref{sec:results} we show the first demonstration of closed-loop on-sky wavefront sensing and control with FAST. We were able to reduce the strength of the speckle field and improve the contrast in the desired region while operating at sub-micron near infrared wavelengths, with a narrowband filter centered at 875 nm and a bandwidth of about 3\%. Finally, sections \ref{sec:conclusions} contains a summary and future steps.

\section{METHOD}\label{sec:method}
\subsection{Optical layout}
\begin{figure}[h]
    \centering
    \includegraphics[width=0.85\linewidth]{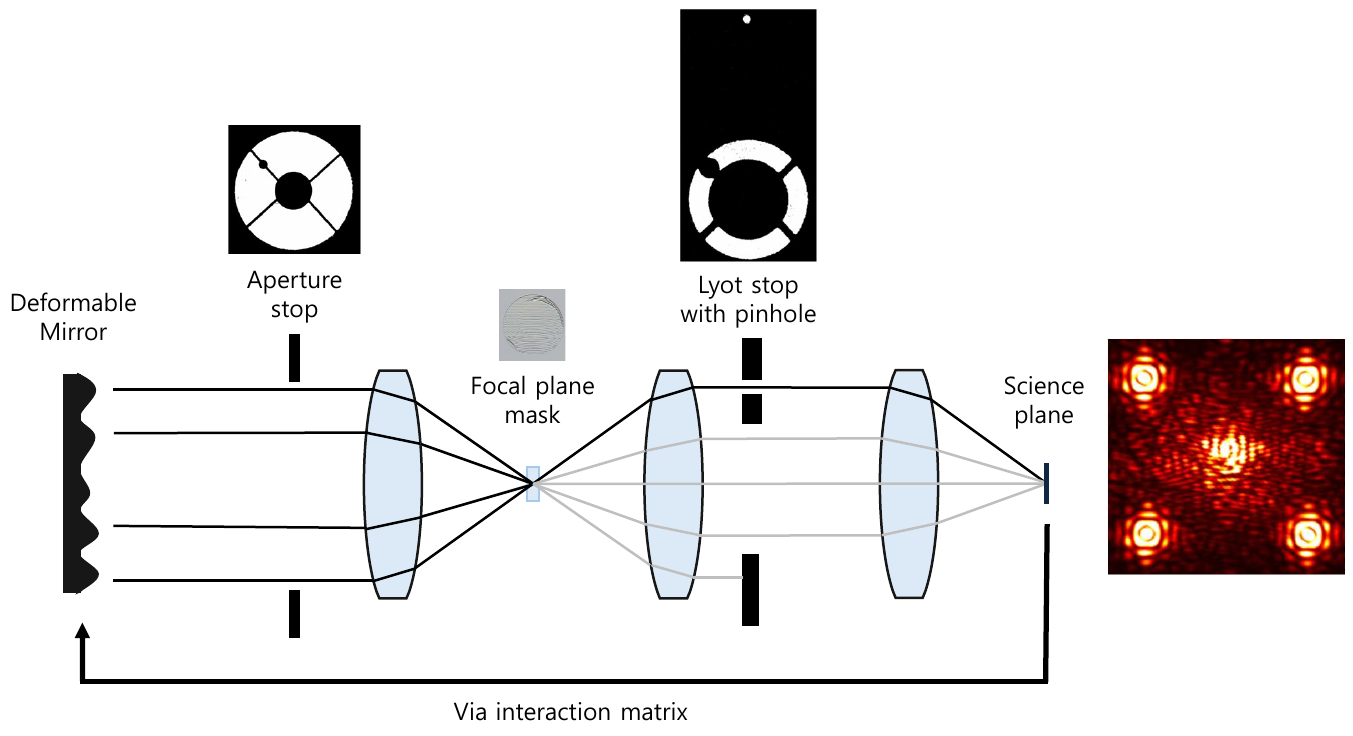}
    \caption{FAST optical layout in MagAO-X. It features, a 1K DM for NCPA correction; MagAO-X’s binary aperture stop called ``bump mask” used for coronagraphic observations; a focal plane phase mask with a wedge, diffracting light at a specific angle; and a Lyot stop with a small pinhole placed at an adequate distance. On the science plane, the on-axis source is fringed because of the recombination and interference of on-axis light leaking through the focal plane mask and the reference beam created by the pinhole. The image shown is a laboratory measurement. A single measurement like this is required for the wavefront sensing. Via the estimation of the stellar complex speckle field and an interaction matrix previously calibrated, the DM can be used to perform wavefront control.}
    \label{fig:layout}
\end{figure}

MagAO-X is the extreme adaptive optics system for the 6.5-meter Magellan Clay Telescope at Las Campanas Observatory in Chile. It operates at visible and sub-micron near infrared wavelengths, delivering high Strehl ratios, high resolution, and high contrast. It features a main extreme adaptive optics system consisting of two DMs, as well as a secondary adaptive optics stage placed in front of the coronagraph. This is a 1K MEMS DM dedicated to the correction of NCPAs. The FAST architecture of MagAO-X is shown in figure \ref{fig:layout}.

The DM is the first element of this subsystem, and it’s used for active control of the NCPAs. We use an aperture stop that we call ``bump mask” for coronagraphic observations to mask a DM area where the surface cannot fully deflect because of a manufacturing issue. Light is focused on the FAST phase mask, which does not block any light, but rather diffracts it outside of the geometrical pupil, on the pinhole. The mask is circular and its wedge angle is about 2.3$^{\circ}$. It was designed for a narrowband filter centered at 875 nm with $\sim$3\% bandwidth and to achieve a deflection angle of about 1.2$^{\circ}$ to reach the pinhole location on the pupil plane. The diameter of the mask is roughly 4 $\lambda$/D. The mask was manufactured in-house at Leiden Observatory using the Nanoscribe Photonic Professional GT machine. This is a micro-3D printer performing additive manufacturing with high-precision using two-photon polymerization. The mask was printed on a 1mm thick fused silica uncoated substrate with the highest possible resolution (63x objective) and is made of the photo-resistant resin IP-Dip. Most of the on-axis light is diffracted towards the pinhole, but part of it still leaks through the mask and makes its way through the Lyot stop to the science plane. After the focal plane mask, the light is re-collimated and passes through a Lyot stop with a pinhole. The light passing through the pinhole creates a spatially filtered reference beam. Light is finally focused an recombined on the science plane where it coherently interferes creating fringes. If an off-axis source is present, it does not fall on the mask and so it is not deflected towards the pinhole and it is not included in the reference beam. Therefore, it does not show fringes and it does not interfere with the on-axis light because of incoherence. Fringe measurements can therefore be exploited to estimate and control the stellar speckle field independent of the companion, which will not be affected in any way.

\subsection{Sensing and control principle}
We use iEFC for the focal plane wavefront control. So, first, an empirical calibration is performed to build an interaction matrix connecting the DM commands with modulated intensity measurements in the science plane. In this case, the intensity measurement is the SCC fringe pattern. To do this, we select the region where we want to improve the contrast (or dig a dark hole) and create Fourier modes corresponding to such region and apply them on the DM. The wavefront sensing is achieved through a single image of the SCC fringes for each mode, through a process called ``side-band extraction", explained below. We then use the push-pull method for each mode to build the interaction matrix.

The side-band extraction requires us to record a single image and Fourier transform it to obtain the Optical Transfer Function (OTF). The OTF can be visualized as the convolution of the Lyot stop and pinhole with itself, so, it consists of a central part and two side lobes, where the side lobes are the interference terms of the SCC (the pinhole convoluted with the central Lyot stop). So, we mask the OTF leaving only one side lobe visible and apply an inverse Fourier transform to obtain an estimate of the stellar complex field. For the analytical solution, see Galicher et al. 2010\cite{SCC_FPWFS_CDI}.

During the control, we perform the side-band extraction in real time to estimate the speckle field that we want to suppress. Via the interaction matrix calibrated previously, we then know the DM pattern needed to create such speckle field, and we apply it with opposite sign, to suppress it. We run this in closed-loop to actively dig a dark hole in the desired region, and try to keep it stable during our observations. Since we have enough light in the reference beam, we can run the loop at the speed of the science camera.

\section{ON-SKY CLOSED-LOOP DEMONSTRATION}\label{sec:results}
We observed the bright spectroscopic binary B star Alpha Virginis (also known as Spica) during the 2025A MagAO-X observing run, on the 21st of April 2025 starting at 03:00 UTC. The observatory telemetry recorded a seeing of about 0.8 arcsec during our observations. Moreover, high winds created a strong atmospheric halo with an oblique direction.

\begin{figure}[h]
    \centering
    \includegraphics[width=0.85\linewidth]{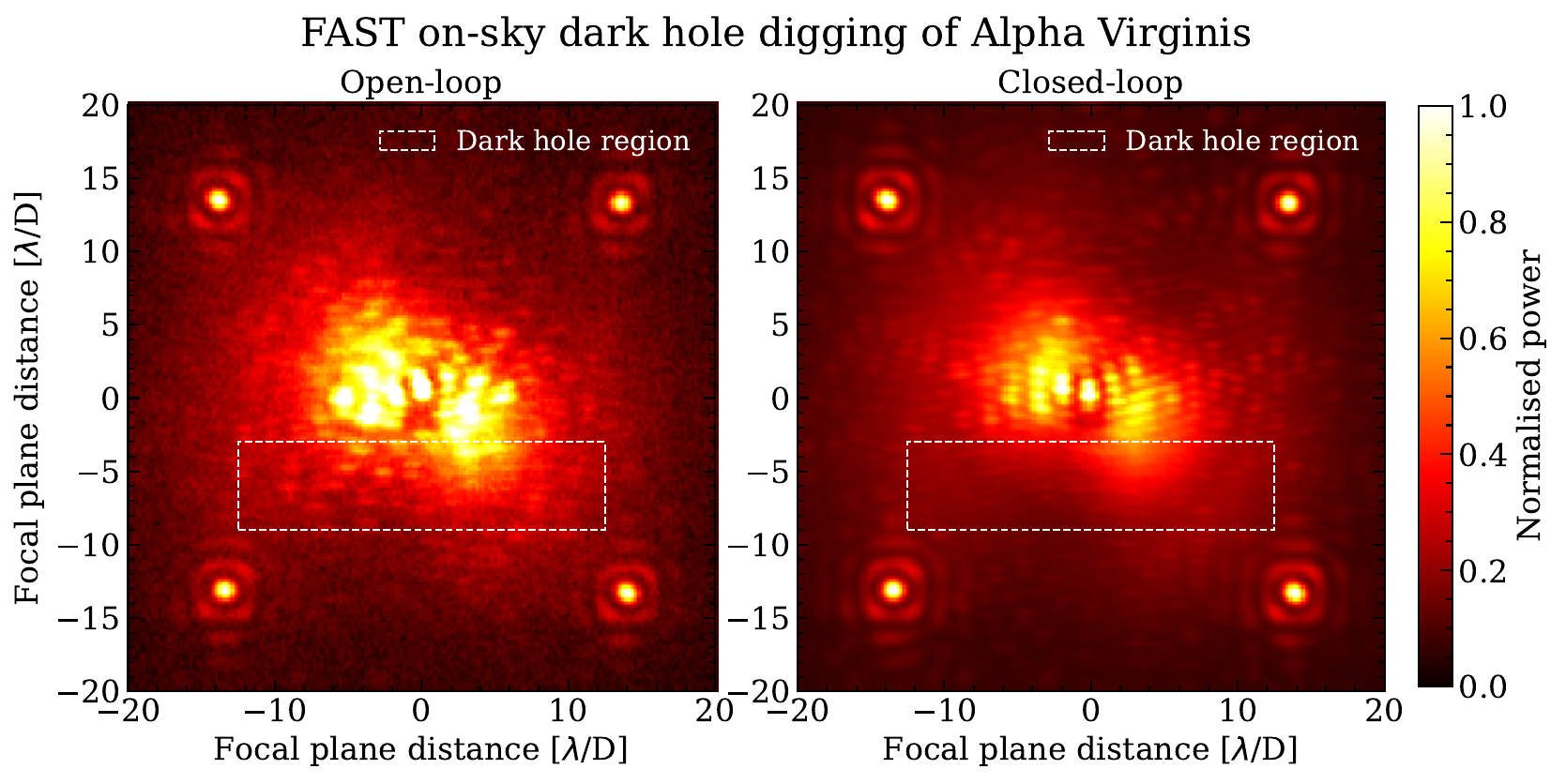}
    \caption{On-sky dark hole digging around Alpha Virginis. The dashed white rectangles delimit the region where we want to improve the contrast. In the left panel no wavefront control is performed, while in the right panel wavefront sensing and control are actively performed on-sky with SCC+iEFC.}
    \label{fig:spica}
\end{figure}

Figure \ref{fig:spica} shows Alpha Virginis observations with and without active wavefront sensing and control with FAST. The dark hole region is located below the central star with an IWA of 3 $\lambda$/D and an outer-working angle (OWA) of 9 $\lambda$/D, and a width of 25 $\lambda$/D. When the SCC in running is closed-loop (right panel) the fringes are greatly reduced inside the dark hole while they are still present in the rest of the image. Note that the remaining fringes look smoother on the right panel than in the left one simply because more frames were averaged together during the control. What we want to convey here, is that only in the dark region the speckles are actually suppressed thanks to the control. We achieve a 1-$\sigma$ raw contrast improvement of a factor 10 inside the dark hole region in closed-loop. This is a good result that shows for the first time a closed-loop SCC working on-sky.

\section{CONCLUSIONS}\label{sec:conclusions}
We have shown preliminary results from the first on-sky active focal plane wavefront sensing and control demonstration with an SCC. We performed this with MagAO-X using a narrowband filtered centered at 875 nm. Thanks to the SCC and FAST principles, we were able to sense the stellar complex field with a single measurement and in real time. The control was performed actively through iEFC. We achieved a factor 10 improvement in 1-$\sigma$ raw contrast with closed-loop control inside the dark hole region. Since the SCC also allows for Coherent Differential Imaging (CDI) in post-processing\cite{SCC_FPWFS_CDI}, the contrast could be improved even further.

In the future, we will show more observations of stars with companions, and approach the chromaticity problem of the phase mask to increase the operational bandwidth of the SCC. The SCC will also benefit of future improvements to the MagAO-X system that are already planned, including the management of bench turbulence and of telescope vibrations with predictive control. Moreover, we want to design a new Lyot stop with higher throughput. Finally, we will create an optical model of the system and perform CDI to retrieve information about companions and improve the contrast even further.

\acknowledgments 
This paper includes data gathered with the 6.5 metre Magellan Telescopes located at Las Campanas Observatory, Chile. The MagAO-X phase II project acknowledges generous support from the Heising-Simons Foundation. We are very grateful for support from the NSF MRI Award \#1625441 (MagAO-X). MagAO-X uses the CACAO software package, which is supported by NSF Award \#2410616. SYH acknowledges support from NWO Award 184.036.004. JL and SYH acknowledge support from NASA APRA award 80NSSC24K0288.

\bibliography{report} 
\bibliographystyle{spiebib} 

\end{document}